\documentclass[useAMS,usenatbib]{mn2e}
\usepackage{epsfig}
\usepackage{times}

\newcommand{\etal}
	{et al.}

\newcommand{\bj}
        {b_{\rm J}}

\begin{document}

\title[Correlating galaxy morphologies and spectra in the 2dFGRS] 
{Correlating galaxy morphologies and spectra in the 2dF Galaxy
Redshift Survey}

\author[D.S Madgwick]    
       {Darren S.\ Madgwick\thanks{
                E-mail: dsm@ast.cam.ac.uk (DSM)} \\ 
        Institute of Astronomy, Madingley Road, Cambridge
        CB3 0HA, U.K.}

\date{
Accepted 2002 September 3. 
Received 2002 June 7; in original form 2002 April 14}

\pagerange{\pageref{firstpage}--\pageref{lastpage}}
\pubyear{2002}

\label{firstpage}

\maketitle

\begin{abstract}
The correlation between a galaxy's morphology and its
observed optical spectrum is investigated.
As an example, 4000 galaxies from the 2dF Galaxy Redshift
Survey, which possess both good quality spectra and have visually
determined morphologies are analysed.  
Of particular use is the separation of Early
and Late type galaxies present in a redshift survey since these can
then be used in their respective redshift-independent distance
estimators ($D_n-\sigma$ and Tully-Fisher).  
It is determined that galaxies in this sample can be relatively
successfully separated into these two types by the use of
various statistical methods.  These methods are briefly outlined in 
this paper
and are also compared to the default 2dFGRS spectral classification $\eta$.   
In addition it is found that the
4000\AA\ break in the spectrum is the best discriminant in
determining its morphological type.
\end{abstract}

\begin{keywords}
methods: statistical
--
galaxies: elliptical and lenticular, cD
--
galaxies: spiral
\end{keywords}

\section{Introduction}
\label{section:intro}

The classification of galaxies according to their observed
morphologies has proved to be a very useful way of characterising
different galaxy populations (see e.g. Hubble 1936).  However, the
high-resolution imaging data required to make such accurate morphological
classifications, over a wide range of redshifts, is often unavailable in
large galaxy redshift surveys. 
In this paper, a selection of different statistical methods
are investigated, with the aim of establishing a
quantifiable link between a galaxy's morphology
and its observed optical spectrum; thousands of which are now 
available through the advent of large galaxy redshift surveys such as the 
Sloan Digital Sky Survey (York \etal\ 2000) and the  
2dF Galaxy Redshift Survey (2dFGRS, Colless et al. 2001).  
In particular, a set of 4000 galaxies which have already been 
observed in the 
2dFGRS, for which the morphology and spectrum have been
determined, is used as a training set.
It has long been established
that a substantial link exists between the overall structure of a galaxy and
the chemical properties reflected in its spectrum - which quantifies
its stellar and gas composition (e.g. Morgan \& Mayall 1957).
However, efforts 
to quantify this relationship have been somewhat hampered due to the lack of
large, representative, data-sets.  For example, Folkes, Lahav \& Maddox (1996)
used a sample of only 26 unique galaxy spectra and morphologies in a
similar analysis to that presented here.
This is fortunately no longer an issue with the
advent of fibre-based galaxy redshift surveys, which are able to acquire 
several hundred galaxy spectra per hour.

These surveys
are producing extremely large data-sets for which the
spectrum of a galaxy is known but its detailed structural parameters
generally are not, 
or are very difficult to
determine accurately from the photometry of the input catalogue over a
representative redshift range.
Fortunately, the spectrum of a galaxy is generally a 
more robust quantity to measure over a variety of redshifts and as such if
a substantive link between optical spectra and these parameters can be determined this 
will greatly enhance our ability to probe the properties of our local 
galaxy population.

Another more specific advantage of being able to determine a galaxy's
morphology from its 
spectrum is that it allows the identification of targets 
for peculiar
velocity follow-ups (using either $D_n-\sigma$ for elliptical galaxies or the
Tully-Fisher relation for spirals).

In this paper two statistical techniques are investigated, to
determine how accurately morphology can be estimated from a galaxy's optical 
spectrum.  Both methods are `supervised' - meaning that they require a
training set of galaxies with both visually determined morphologies
and observed spectra.
The first method to be implemented is Fisher's linear discriminant
(Section~4),   
which attempts
to determine an optimal linear combination of inputs (the spectrum) to 
distinguish between several outputs (the morphologies).  The second
method is an Artificial Neural Network (ANN, Section~5) which creates 
non-linear combinations of the input, and outputs a selection of class
probabilities.  The possible biases introduced into this work 
by systematic effects are described in Section~6 and then the success
rates of each method are compared to those achieved using the default
2dFGRS spectral classification parameter, $\eta$, in Section~7.

\section{Galaxy Morphologies in the 2dFGRS}
\label{section:section2}

\begin{figure*}
\epsfig{file=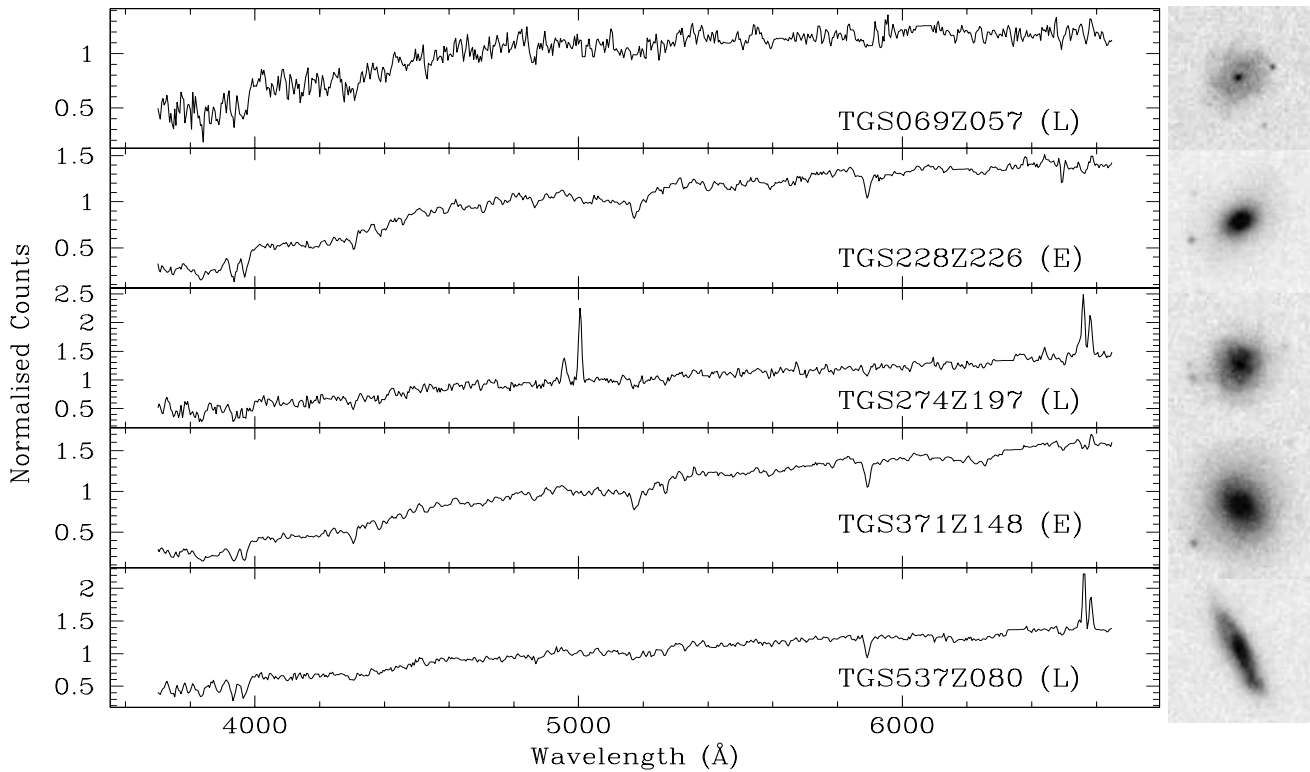,width=7in}
\caption{A selection of example galaxy spectra and images are shown.
The spectra have been taken from the 2dF Galaxy Redshift Survey and
are in units of counts/bin with arbitrary normalisation.  The images
shown are $1\times 1$ arcmin postage stamps, in the standard
astronomical orientation, taken from the SuperCOSMOS 
Sky Survey (Hambly \etal\ 2001).  For each spectrum the 2dFGRS object
name is given and the labels (L) and (E) refer to Late-type and Early-type
morphologies respectively.}
\label{fig:exspec}
\end{figure*}

A number of galaxies in the 2dFGRS have already had their morphologies
determined manually by direct examination of the APM Galaxy Survey
images (Maddox \etal\ 1990a,b see Fig.~\ref{fig:exspec}).  
The accuracy and completeness of this
sample varies a great deal depending on the source of the
classification and the range of galaxy magnitudes considered.  In what
follows I primarily make use of those galaxies which were ascribed morphologies
in the APM Bright Galaxy Catalogue (Loveday 1996). This catalogue provides a
complete sample of classified galaxies down to a magnitude limit of 
$\bj=16.44$.  The exact value of this magnitude limit now varies
across the sky due to recent re-calibrations of the APM magnitudes
(see e.g. Colless \etal\ 2001 for details of the most recent
calibrations of those galaxies included in the 2dFGRS), however this
will not have any substantial impact upon the representativeness of
our classified sample.  A significant number of galaxies at fainter
magnitudes also have morphologies determined from other sources, 
but these will be
substantially less reliable and so are not used in this analysis.

\begin{figure}
\epsfig{file=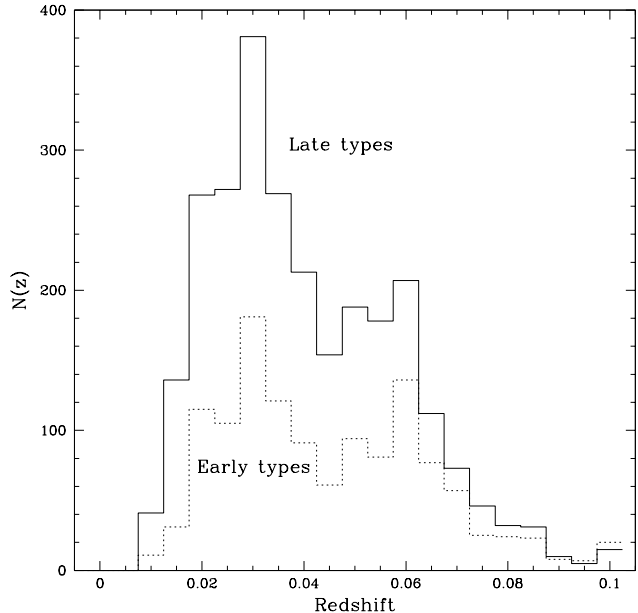,width=3.5in}
\caption{The distribution of redshifts of those galaxies classified as
Early or Late in the 2dFGRS.  Note that in order to ensure that the
classification is relatively robust we only include those galaxies
within the magnitude limit $\bj<16.5$.}
\label{fig:nofz}
\end{figure}

Of those $\bj<16.5$ galaxies which have been successfully observed so
far in the 
2dFGRS, 3899 have a morphological classification (Fig.~\ref{fig:nofz}).
The galaxy morphologies are given in four broad bins: Elliptical, 
S0, Spiral and
Irregular.  However, in the analysis presented here they are rebinned
into only two classes; Early (Elliptical,S0) and
Late (Spiral,Irregular) types.  The reasons for doing so are two-fold:
Firstly, the number of classified galaxies which have been identified as
S0 or Irregular are significantly smaller than the number of Spirals.  
Therefore
the identification of these types will be greatly hindered by the
presence of Spiral outliers - which effectively swamp out any identifying
signal which may arise from these types.  Secondly, the distinction between
Early and Late type galaxies is of fundamental importance to observational
cosmology since each can be used in its own redshift-independent distance
estimator, i.e.  $D_n-\sigma$ for Early types (Dressler \etal\ 1987)
and the Tully-Fisher relation for Late types (Tully \& Fisher 1977).

Note that this sample of galaxies consists entirely of relatively bright,
nearby galaxies and so may not be representative of the entire 2dFGRS
galaxy population.
Another important point to bear in mind is that as
these galaxies are relatively extended on the sky, 
the spectra observed of 
them (through the fixed fibre aperture of the 2dF instrument) 
may not be representative of the entire galaxy.  This so called 
`aperture effect' is a difficult issue to address and has led to much
discussion in the literature (see e.g. Kochanek, Pahre \& Falco, 2000;
Madgwick \etal\ 2002).  The possible impact of aperture effects on our
results is discussed further in 
Section~\ref{section:aper}.

\section{Galaxy spectra in the 2dFGRS}

\begin{figure*}
 \epsfig{file=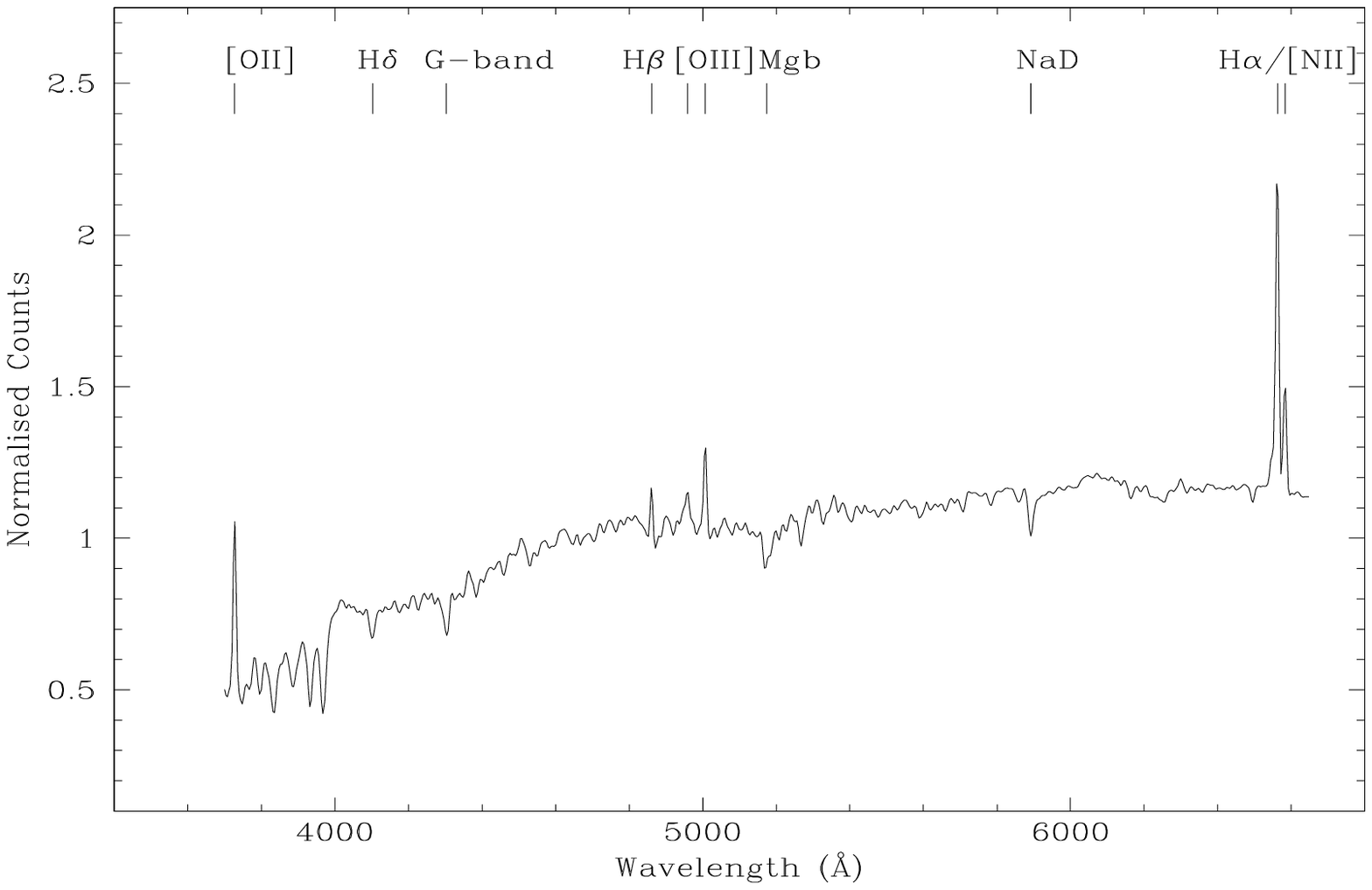,width=6in}
 \caption{The average spectrum of an $M_{\bj}-5\log_{10}(h)>-18$
 volume-limited sample of galaxies drawn from the 2dF Galaxy Redshift
 Survey.   The main spectral features
 present are labelled.}
 \label{fig:aver}
\end{figure*}

The purpose of this analysis is to relate the spectrum of a galaxy to
its morphology.  In the 
case of the 2dFGRS each spectrum consists of 1024 channels spanning
the wavelength range of approximately 3700-8000\AA, thereby including
all the major optical diagnostics between O{\sc [ii]} and H$\alpha$
(see Folkes \etal\ 1999, for further details).  As an illustration,
the average 2dFGRS spectrum for a representative volume-limited sample
is shown in Fig.~\ref{fig:aver}.

Rather than dealing with all 1024 spectral channels in the subsequent
analysis (to represent a given spectrum),
it is possible to take advantage of the fact that the vast majority of
these channels 
are redundant by means of some form of data compression.  In the
analysis presented here use is made of a Principal Component Analysis (PCA, 
see e.g. Murtagh \& Heck 1987) since this compression algorithm has met with
considerable success in dealing with galaxy spectra
(e.g. Connolly \etal\ 1995; Galaz \& de Lapparent 1998; Folkes \etal\ 1999; Madgwick \etal\ 2002).

\subsection{Review of PCA}

PCA is a well established statistical technique which has proved
very useful in dealing with high dimensional data sets.  In the
particular case of galaxy spectra we are typically presented with
approximately
1000 spectral channels per galaxy, however when used in applications this is
 usually compressed down to just a
few numbers, either by integrating over small line features - yielding
equivalent
widths - or over wide colour filters.  The key advantage of using PCA in
our data compression is
that it allows us to make use of all the information contained in the spectrum
of a galaxy in a statistically unbiased way, i.e. without the use of
such ad hoc filters. 
 
In order to perform the PCA on our galaxy spectra we first construct a
representative volume limited sample of the galaxies.  When we apply
the PCA to
this sample it constructs an orthogonal set of components (eigenspectra, 
herein denoted ${\mathbf{PC}}_1$,${\mathbf{PC}}_2$,etc)
which span the wavelength space occupied by the galaxy spectra.  These
components have been specifically chosen by the PCA in such a way that as much
information (variance) is contained in the first eigenspectrum as possible,
and that the amount of the remaining information in all the subsequent
eigenspectra is likewise maximised.  Therefore, if the information contained
in the first $n$ eigenspectra is found to be significantly greater than that
in the remaining eigenspectra we can significantly compress the data set
by swapping each galaxy spectrum (described by 1000 channels) with
just those first $n$ projections (denoted $pc_1$,$pc_2$ etc).  
The
variances corresponding to the first 10 principal components derived
in this manner are shown in Table~\ref{vari}.

Note that the PCA is merely a statistical tool, we do not imply (yet)
that any of these components are physically significant, but rather
we are merely using them as a method of data compression.
 
\begin{table}
 \caption{Variance contained in the first 10 principal components of
the 2dFGRS galaxy spectra.}
 \begin{tabular}{@{}cccc@{}}
   \hline
   Component & Variance (\%) & Component & Variance (\%)\\
   \hline
     1 & 54 & 6 & 0.99 \\
     2 & 15 & 7 & 0.86 \\
     3 & 4.0 & 8 & 0.57 \\
     4 & 2.7 & 9 & 0.41 \\
     5 & 1.2 & 10 & 0.27 \\
   \hline
 \end{tabular}
 \label{vari}
\end{table}

During the PCA analysis it was found that 
the eigenspectra became dominated by unphysical broad features
from the fifth eigenspectrum (${\mathbf{PC}}_5$) onwards.  This was due to 
artifacts from sky emission features
which we were unable to completely remove during the spectral reduction.
Rather than restrict our future analysis to only the first four 
principal components, we have instead repeated the analysis with the
wavelength range 5850-6200\AA\ masked out.  This has no noticeable
effect on the original first four eigenspectra, all of which are essentially
identical between the two PCA implementations.  However, 
it does allow us to probe deeper
into the PCA space than we would otherwise have been 
able to (up to ${\mathbf{PC}}_9$) .

It is worth noting that most galaxy spectra can be accurately reconstructed
using only the first 4-5 principal components.  In our subsequent 
analysis we make use of the first nine (this choice is justified in
detail by
Folkes, Lahav \& Maddox 1996).

\begin{figure*}
\epsfig{file=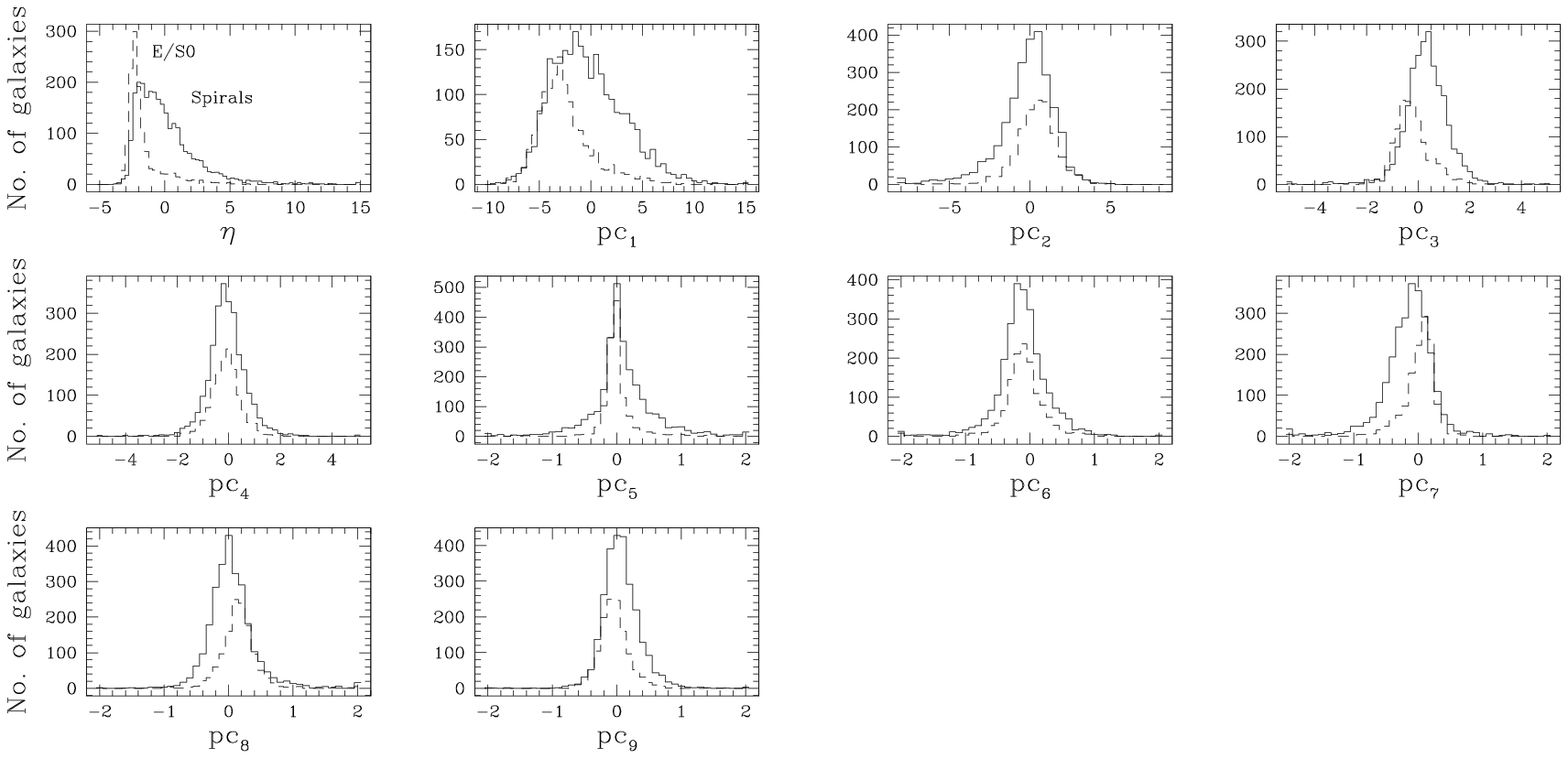,width=7in}
\caption{The distribution of $\eta$ for E/S0 (dotted line) and Spiral 
galaxies (solid line)
is shown in the top left panel.  The other panels show these distributions
for the first 9 principal components.}
\label{fig:pcs}
\end{figure*}

\begin{figure*}
\epsfig{file=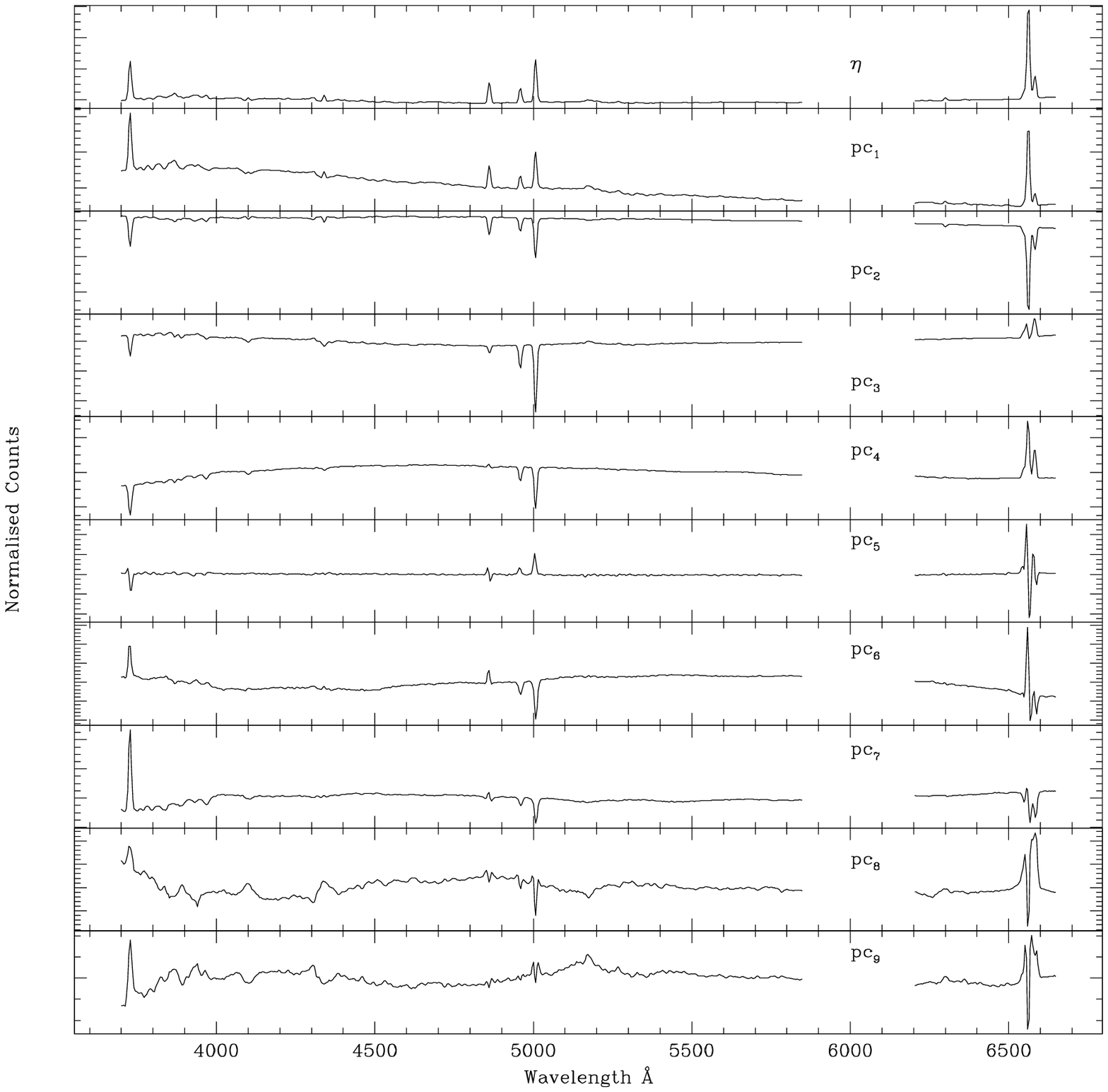,width=7in}
\caption{The first 9 principal components.  Also shown is the spectrum
of the $\eta$ component (top panel), used to classify the 2dFGRS.}
\label{evecs}
\end{figure*}

In order to gain some initial understanding of the link between galaxy spectra
and morphologies, histograms are shown in Fig.~\ref{fig:pcs} 
of the distributions of principal component
projections for both Early and Late type galaxies.
Also plotted are
the eigenspectra for each of these projections, which illustrate what
spectral features are encoded in the respective projections
(Fig.~\ref{evecs}).   
Included in each of these
plots are the corresponding results for the $\eta=0.5pc_1 - pc_2$
projection which is used to spectrally classify the 2dFGRS galaxies
(see next Section).  It can be seen that most of the first nine
principal components encode some degree of information about the
morphology of a galaxy, although none are capable of separating Early
from Late type galaxies accurately (with the exception of the $\eta$
spectral classification).  The aim of the analysis presented in this
paper is essentially to determine if it is possible to improve the
separation between Early and Late types by taking either linear
(Fisher Analysis, Section~\ref{section:fisher}) or
non-linear (Artificial Neural Networks, Section~\ref{section:ann}) 
combinations of these projections.

\subsection{Spectral classification in the 2dFGRS}
\label{sub:class}
 
The 2dF instrument (Lewis \etal\ 2002) was designed to measure large numbers of
redshifts in as short an observing time as possible.
However, in order to
optimise the number of
redshifts that can be measured in a given period of time, compromises
have had to be made with respect to the spectral quality of the
observations.  Therefore if one wishes to characterise the observed
galaxy population in terms of their spectral properties care must be
taken in order to ensure that these properties are robust to the
instrumental uncertainties.
 
\begin{figure}
 \epsfig{file=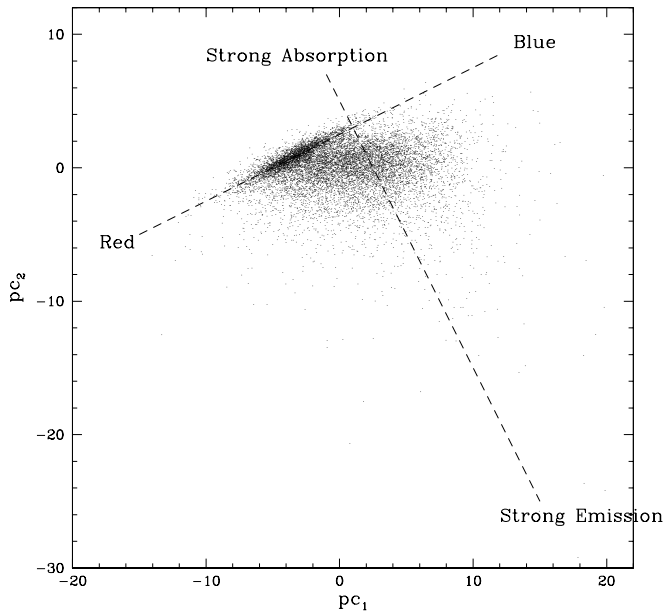,width=3.5in}
 \caption{The distribution of ($pc_1$,$pc_2$) projections of the observed 2dF
 galaxies is shown.  Also shown are the projections which maximise
 either the continuum or emission/absorption component in each
 spectrum.  The latter of these projections is used to classify the
 2dFGRS galaxies.}
 \label{fig:eta}
\end{figure}

The quality and representativeness of the observed
spectra can be compromised in several ways and a full list is
presented in previous work (see e.g. Madgwick \etal\ 2002).  
The net effect is that
the uncertainties introduced into the fibre-spectra
predominantly affect the calibration of the continuum slope and have
relatively little impact on smaller-scale features such as the
emission/absorption line strengths. 
For this reason any given galaxy spectrum which is projected into the
plane defined by ($pc_1$,$pc_2$) will not be uniquely defined in the
direction of varying continuum but will be robust in the orthogonal
direction (which measures the average line strength, see
Fig.~\ref{fig:eta}).

The projection onto this robust axis is denoted by $\eta$ ({\tt
ETA\_TYPE} in the 
2dFGRS catalogue\footnote{\tt http://www.mso.anu.edu/2dFGRS/}),
\begin{equation}
 \eta = a\cdot pc_1 - pc_2 \;.
\end{equation}
Where $a$ is a constant which we find empirically to be $a=0.5\pm
0.1$.  This (continuous) variable $\eta$, being the single
most important component of the galaxy spectra which was robust to
instrumental  uncertainties, was chosen as the measure of spectral
type in 2dFGRS. 

Note that although $\eta$
is a continuous measure of type, it is often useful to divide a galaxy
sample into different bins to simplify subsequent analyses.  One of the
most common divisions to make is to separate those galaxies with
$\eta<-1.4$ from those with $\eta\ge -1.4$, referred to as {\em
relatively} quiescent and star-forming galaxies respectively.

\section{Fisher's Linear Discriminant}
\label{section:fisher}

It is clear from the trends in Fig.~\ref{fig:pcs} that morphology is
indeed 
related to the spectra of galaxies although, with the exception of the 
$\eta$ spectral classification,
the correspondence is not particularly pronounced.  However, as shown
by $\eta$, it may be possible to improve this situation by taking
various linear combinations of the projections in order to isolate those
particular parts of a galaxy's spectrum which contain the most information 
regarding its morphology.
 
In this section a method first proposed by Fisher (see
e.g. Bishop 1995 and references therein) is considered, for linearly
reducing
data dimensionality (in this case we hope to go from our 9 principal
components to 1 morphology)
in a way which will optimally distinguish between different classes of
objects.
Strictly speaking this method will not create a discriminant between the
two classes but rather, once we have reduced the dimensionality, it
should become clear how to divide the sample.

As mentioned before, this
method is a linear approach to this problem.  In the next section this will 
be generalised to consider the possibility of a non-linear relationship
between morphology and spectra through the use of 
Artificial Neural Networks.

\subsection{Mathematical formulation}

The mathematical formulation of Fisher's linear discriminant is 
straight-forward.  We consider a set of input vectors ${\mathbf x}$; we wish 
to determine the weights
$\mathbf{w}$, such that the projection $\mathbf{y}$ is the most
discriminatory between our two classes,
\begin{equation}
 {\mathbf{y}} = {\mathbf{w}}^T{\mathbf{x}}
\end{equation}
The mean vectors ${\mathbf{m}}_1$ and ${\mathbf{m}}_2$ of each class are given by,
\begin{equation}
{\mathbf{m}}_k = \frac{1}{N_k} \sum_{n\in C_k} {\mathbf{x}}^n \;.
\end{equation}
Here $n\in C_k$ are the elements of class $C_k$ (in this case $k=\{1,2\}$ for
Early and Late types respectively).
After projection,
the scalar separation between the means will be,
\begin{equation}
m_2-m_1 = {\mathbf{w}}^T ({\mathbf{m}}_2-{\mathbf{m}}_1) \;,
\end{equation}
and the {\em within-class} covariance is,
\begin{equation}
s^2 = \sum_{n\in C_k} (y^n - m_k)^2
\end{equation}
Fisher's assertion was that the optimal mapping to be used in reducing
our dimensionality should be such as to maximise,
\begin{equation}
J({\mathbf{w}}) = \frac{(m_2 - m_1)^2}{s_1^2 + s_2^2}\;.
\label{eqn1}
\end{equation}
This is easy to interpret physically:  We wish to find a set of weights
$\mathbf{w}$ such as to maximise the mean separation between the two
classes $(m_2-m_1)$ after we have performed our projection.  The
variance in the denominator takes into account the fact that our
initial vectors $\mathbf{x}$ may have different spreads and hence differing
degrees of overlap in different directions.
Clearly we wish to find a projection which minimises
this overlapping between our two classes.
 
To determine the optimal weights we simply need to re-introduce the
explicit weight dependence into Eqn.~\ref{eqn1} and differentiate.  It
is shown 
in Bishop (1995) that the solution is then,
\begin{equation}
{\mathbf{w}} \propto {\mathbf{S_w}}^{-1} ({\mathbf{m}}_2-{\mathbf{m}}_1) \;.
\label{eqn2}
\end{equation}
Where,
\begin{equation}
{\mathbf{S_w}} = \sum_{C} \sum_{n\in C} ({\mathbf{x}}^n-{\mathbf{m}}_c)({\mathbf{x}}^n-{\mathbf{m}}_c)^T
\end{equation}
is the total {\em within-class} covariance matrix.  Equation~\ref{eqn2}
only yields the direction of the weight vector, not its
magnitude.  Conventionally the magnitude of $\mathbf{w}$ is taken to be such
that $\sum_i w_i^2 = 1$.

\begin{figure}
\epsfig{file=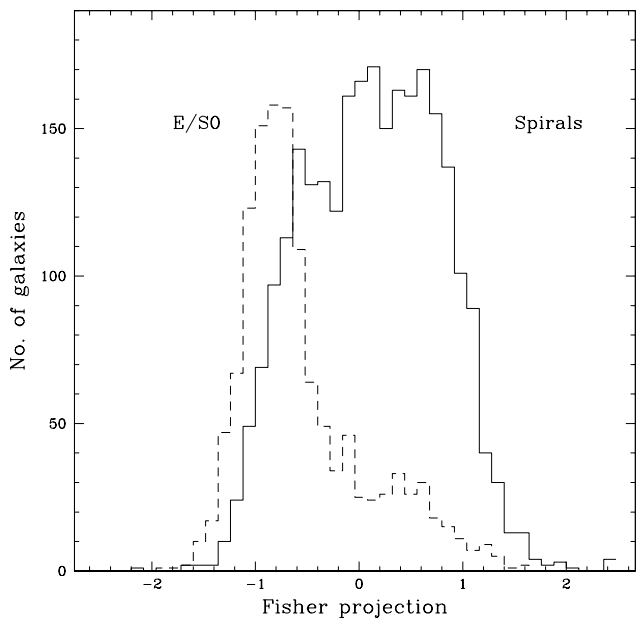,width=3.5in}
\caption{Fisher's linear discriminant, as calculated from the first nine
principal components with the aim of discerning the galaxy morphology.
It is clear a significant degree of overlap between the morphological
types still exists.}
\label{fisher1}
\end{figure}

\begin{figure*}
\epsfig{file=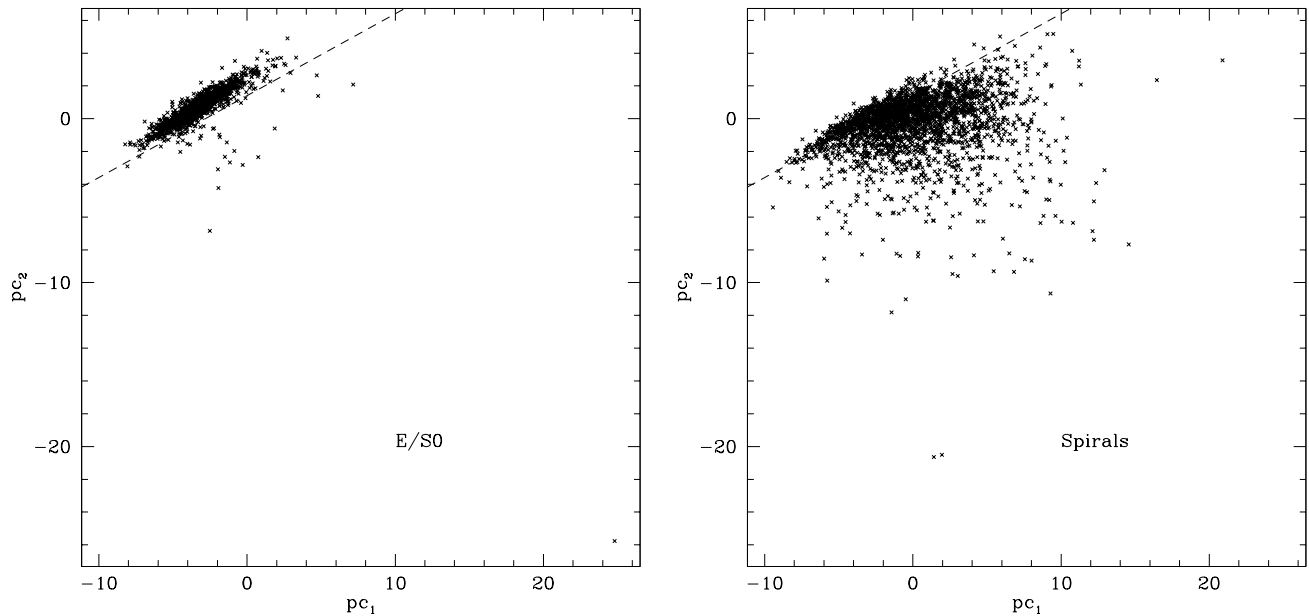,width=7in}
\caption{The $pc_1$ and $pc_2$ projections of the morphologically 
classified 2dFGRS galaxies are shown.  Here the morphologies have been
derived using Fisher's method with 9 principal components.  It is remarkable
that all the information in these 9 components is essentially contained
in these first two projections.  The Fisher discriminant has been cut
at -0.55.}
\label{fisher2}
\end{figure*}

\subsection{Classification using the Fisher analysis}
\label{section:fres}

After performing the Fisher analysis on the complete set of morphologically
classified 2dFGRS galaxies,
the resulting weighting vector is found to be,
\begin{equation}
  {\mathbf{w}} = (1.2,-1.3,3.5,0.18,0.34,-0.96,-0.68,-6.5,6.4)/10 \;.
\end{equation}
Figure~\ref{fisher1} shows the resulting Fisher projections 
using these weights for both the known Early and Late type galaxies in
the 2dFGRS.    It is clear from this
figure that the method has been relatively successful in that the
distinction between Early and Late type galaxies is now much more pronounced
than for any of the individual principal components (Fig.~\ref{fig:pcs}).

Having derived this projection it is now necessary to determine `by-hand'
a cut that we will use to distinguish Early from Late type galaxies in
the future.  The actual cut adopted will, of course, depend on the specific
application for which one requires the galaxies to be classified.  For
example, if one requires a
complete sample of Early type galaxies
a relatively high cut should be adopted compared to if one requires a
relatively `pure' sample.  We note that in the context of deriving
a pure sample, the results appear to be
somewhat disappointing in that the Fisher analysis
does not
appear to be as discriminatory between Early and Late type galaxies as
would otherwise have been hoped.

From Fig.~\ref{fisher1} we conclude that a good general 
distinction between Early and Late type galaxies can be achieved by
cutting the distribution at $-0.55$.
Using this cut
gives the total number of galaxies classed Early and Late as
(814,454) and (475,2156) respectively, where ($x1$,$x2$) represents
$x1$ galaxies that are genuinely Early type and $x2$ that are
genuinely Late type.  Note that the contamination between types will
vary substantially depending on the selection criteria of the redshift
survey, and so we are primarily interested in the success rates of the
classification i.e. 63\% of Early types and 82\% of Late types
successfully classified, as
opposed to the degree of contamination. 

The distribution shown if Fig.~\ref{fisher1} is very similar 
to that shown in Fig.~\ref{fig:pcs}
for $\eta$.  Indeed, Fig.~\ref{fisher2} shows
the $(pc_1,pc_2)$ projections
for our sample of galaxies separately, depending on whether they
satisfied the Fisher $-0.55$ criterion for being Early or Late type.  
Over-plotted
on this figure is the cut imposed to classify the 2dFGRS into
Type 1 (Quiescent) galaxies using $\eta$ (see Section~\ref{sub:class}).  
The correspondence is remarkable.
Clearly much of the physical information carried in the PCA
eigenspectra is contained in the first two principal components.
It is also interesting that repeating the Fisher analysis 
using only these first two principal
components yields a classification parameter very similar to 
$\eta$ (with very similar success of classification).  These two forms
of classification (Fisher and $\eta$) are contrasted further in
Section~\ref{section:specclass}. 

It is worth noting that most spectral classifications are in fact
indirectly anchored to morphological types by use of a training set
(e.g. the Kennicutt Atlas, Kennicutt 1992) of galaxy spectra for which
the morphology is 
known (see e.g. Connolly \etal\ 1995; Folkes \etal\ 1999).  
However, in the case of the 2dFGRS, $\eta$ was chosen purely on
the grounds of robustness with respect to instrumental uncertainties in
determining the spectral continuum.  This point is discussed further in
Section~\ref{section:specclass}.

\subsection{Spectral features}

\begin{figure*}
\epsfig{file=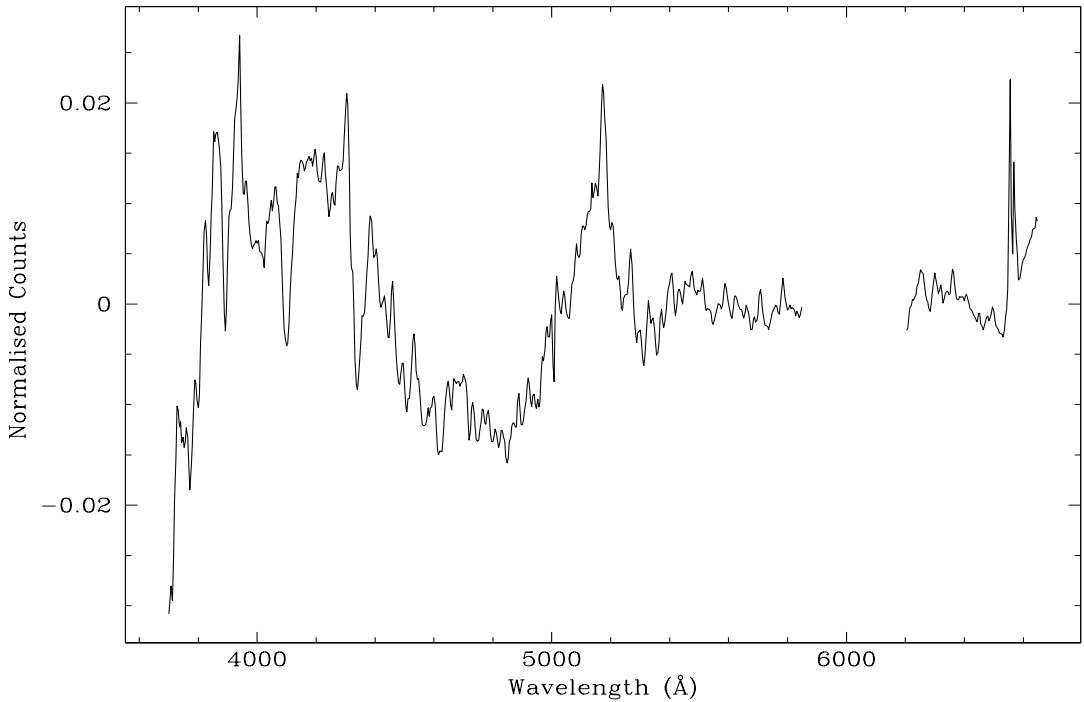,width=6in}
\caption{The combined 2dFGRS eigenspectrum corresponding to the Fisher
criterion derived for determining galaxy morphology.  It can be shown
that the 
Fisher method is almost exclusively quantifying the magnitude of the
4000\AA\ break, and that the entire classification can be created
using only the part of this spectrum at wavelengths below 4800\AA. 
The broad feature at $\sim$5200\AA\ (corresponding to Mgb)
is erroneous, and has resulted from the fact that we have
truncated the series of eigenspectra just as this was feature was
becoming prominent.} 
\label{spec}
\end{figure*}

If we combine the PCA eigenspectra (Fig.~\ref{evecs}) using our
derived weighting vector  $\mathbf{w}$
we can determine which spectral features are being used by the Fisher
criterion to measure galaxy morphology.  This spectrum is shown in
Fig.~\ref{spec}.

It must be borne in mind, when attempting to interpret this spectrum,
that the 
PCA is mean-subtracted, i.e. the average 2dFGRS spectrum has been subtracted
from each individual spectrum before calculating it's projection (see
Fig.~\ref{fig:aver}).  Therefore 
negative values in this eigenspectrum represent either absorption or below
average emission, whereas positive values correspond to above average
emission. 

Perhaps the most striking feature of the spectrum shown in
Fig.~\ref{spec} is its overall lack of nebular emission features,
despite the fact that these dominate most of the PCA eigenspectra
(see Fig.~\ref{evecs}).
It is interesting that this spectrum, which is so clearly different to that
used to calculate $\eta$ (see top panel of Fig.~\ref{evecs}), 
should give such a similar classification.
Clearly there is a very pronounced correlation between the nebular emission
features used to calculate $\eta$, and the strength of the 4000\AA\ break 
and (to a lesser degree) the H$\alpha$ emission line, as measured by
this combined eigenspectrum.  
The apparent broad feature at 5200\AA\ (corresponding to the Mgb
absorption line) is an artifact introduced due to 
our truncating the series of eigenspectra just as this feature was 
becoming prominent in them (it can be clearly seen in ${\mathbf{PC}}_8$ 
and ${\mathbf{PC}}_9$ in Fig.~\ref{evecs}), and as such this
is not contributing to the success of the
classification itself.
 
In fact it is possible to determine more accurately exactly what parts
of this spectrum are being used by the Fisher criterion to determine
morphology by
`masking-out' certain segments of the combined eigenspectrum of
Fig.~\ref{spec}, 
and then seeing how accurately one can recover the same
classification (after re-projecting the galaxy spectra onto it).
By doing this it can be shown that the entire classification is based
on the part of the eigenspectrum below 4800\AA, and hence that the
morphology of a galaxy can be most accurately determined from the
size of the 4000\AA\ break, without the use of any further spectral
information (although using the entire spectrum does not degrade the
result, and hence will continue to be used throughout this work). 
 
\section{Artificial Neural Networks}
\label{section:ann}

This section further refines the analysis of the previous section by
considering the possibility 
of a nonlinear relationship between a galaxy's spectrum and its morphology.
This is done by  making use of an Artificial Neural Network, 
which is trained to identify galaxy morphology after being inputted
with the galaxy spectra (from the first nine PCA eigenspectra).

\subsection{Brief description}

An artificial Neural Network (ANN) is a mathematical construct originally
designed to simulate the functioning of the brain (see e.g. Ripley
1996).  
The network itself
is designed around a set of layers, consisting of input nodes, hidden nodes
and output nodes.
The connections between these nodes can be quite complex and in most
instances all the nodes in a previous layer are connected to all the
nodes in the next layer.  Because of this complexity the ANN can train
itself to recognise highly nonlinear relationships between the inputs and
the outputs we desire.

Each connection between two nodes has an associated weight, $u_{ij}$.  The
total input to any given node is the sum of all the individual inputs
weighted by $u_{ij}$.  Each node takes this total input and applies a
transfer function to it, before passing it on to the next node.  The
transfer function is generally taken to be a sigmoid function,
\begin{equation}
f(z) = \frac{1}{1+\exp(-z)} \;,
\end{equation}
since this keeps the output in the range [0,1], hence limiting the variation
in the possible weighting schemes.
 
In order to train the network we need to establish a cost function.  This
is usually taken to be the Euclidean separation between the desired outputs,
$T_i$, we hope to receive (at each of our output nodes), and the actual
outputs, $F(\mathbf{u},x_i)$,
\begin{equation}
E = \frac{1}{2}\sum_{i=1}^N [T_i - F({\mathbf{u}},x_i)]^2
\label{eqn:error}
\end{equation}
All the weights in our network are initially set to random values.  The
network is then trained on our training data-set by inputing the first four
principal components and comparing the output to the desired morphology of each
object.  The weights are adjusted by means of back-propagation until
a global minimum in the cost function is found.  The network is then
fully trained and can be run on the testing set to establish its
success.

Another aspect of neural networks is that of weight decay, whereby one
can specify an additional factor in the cost function in order to
restrain the magnitude of the weights.  We do not make use of this refinement
for the following reason:  If one specifies multiple outputs for the neural
network (in our case two outputs, corresponding to Early and Late) these
can in fact be interpreted as the
probabilities that a galaxy belongs to either class - so long as the
weight decay is set to 0.

Further details about the mathematical formulation and interpretation
of ANNs is given in e.g. Lahav \etal\ (1996) and Bishop (1995).

\subsection{Results from the ANN}

The ANN was run using three different configurations, all of which had
9 inputs and 2 outputs, and as such were only distinguished 
by the structure and size of their hidden layers.  The first network
used had a single hidden layer of 5 nodes (configuration 9:5:2), 
the second increased this to
9 nodes (configuration 9:9:2), and the final ANN had two hidden layers
with 5 nodes each (configuration 9:5:5:2). 

In each case the network required both a `training' and a 
`testing' data set.  Because there were so many more Late than Early type
galaxies in our sample (2631 versus 1268 respectively),
800 of each type were randomly selected for the purposes of training.
The reason that equal 
numbers of each type were used for training the ANN was because ANNs are
essentially complicated Bayesian classifiers, as such the 
detection of each type will be biased by
by the prior distributional information of that type, which is determined by the
selection criteria of the redshift survey.  By using equal numbers of
each type of galaxy in our training set we ensure that the results
presented here are general and not only applicable to $\bj$-selected galaxy
samples, which are biased towards recently star-forming galaxies.

The remaining 2299 galaxies were used as a testing and validating set.
The purpose of validation is to ensure that the ANN does not `over-train' on
the training set by over-fitting the data which it is supplied.  For this
reason the minimum in the error function (Eqn.~\ref{eqn:error}) is calculated
from the testing data set, rather than the training set.
Note that for our purposes a galaxy was classified as Early type if the
output probability from the ANN from the Early type node was greater than 
0.5, and likewise for Late types.

In general it was found that the results were relatively insensitive to the 
choice of network architecture (changing by only a few percentage points of
successfully detected galaxies of each type).  
This is consistent with the results from
the previous section where it was found that the PCA eigenspectra
generally carry the same physical information as each other - despite being
`statistically' independent.

\begin{figure*}
\epsfig{file=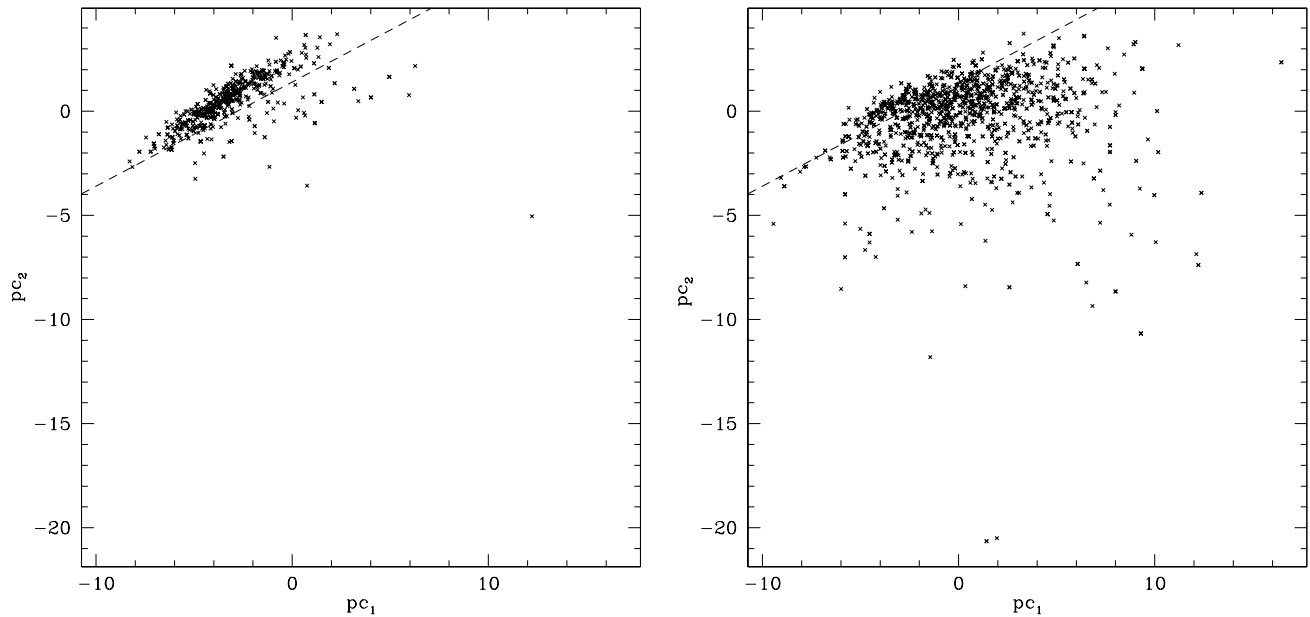,width=7in}
\caption{The $pc_1$ and $pc_2$ projections of the morphologically 
classified 2dFGRS galaxies are shown.  Here the morphologies have been
derived using an Artificial Neural Network with 9 principal components as
inputs (9:9:2).  Again we can see that 
all the information in these 9 components is essentially contained
in just these first two projections.}
\label{fig:net}
\end{figure*}

The results from each ANN are summarised in Table.~\ref{ann}, and
once again the ($pc_1$,$pc_2$)
components are shown for the galaxies which have been classified by the
ANN (in this case the 9:9:2 configuration) in Fig.~\ref{fig:net}.  
Again it can be seen that the classification can be quite accurately
expressed using only these first two projections.  
The correspondence with the 2dFGRS $\eta$ classification is also shown.

One problem that arose when attempting to separate the Early type galaxies 
in our sample was that we obtained a significant contamination from
Late type galaxies ($\sim50$\%).  It was found that this 
fractional contamination could be reduced somewhat by increasing the 
complexity of the ANN, however this resulted in a lower percentage of the
actual Early type galaxies being correctly classified.  
The actual degree of this contamination will vary according to the
galaxy population under consideration.  In the case of the 2dFGRS we
are presented with a $\bj$-selected sample of galaxies and as such there is a
significantly higher proportion of Late type (star-forming) galaxies than Early
types.  On the other hand, in the case of a 2MASS (Jarrett \etal\
2000) near-infrared selected galaxy sample, the proportion of types is
reversed (Madgwick \& Lahav 2001) and so a more pure sample of Early
types can be determined.

\begin{table}
 \caption{Success rates of the ANNs.  The numbers given are
  percentages of galaxies correctly classified for each morphological
  type.}
 \begin{tabular}{@{}cccc@{}}
   \hline
   Network & Configuration & Early & Late  \\
   \hline
     1 & 9:5:2   & 67\% & 80\% \\
     2 & 9:9:2   & 70\% & 82\% \\
     3 & 9:5:5:2 & 67\% & 83\% \\
   \hline
 \end{tabular}
 \label{ann}
\end{table}

\section{Aperture effects}
\label{section:aper}

Because our sample of morphologically classified galaxies is drawn from
only the most nearby (and hence the most extended on the sky)
galaxies,  we must
consider the possibility that the observed spectra are not representative
of the galaxies as a whole.  For example, the 2dF fibre (diameter
2-2.16$''$) may only sample
the light from the bulge of a nearby spiral galaxy - the spectrum from 
which tends to be more similar to that of an Early type galaxy.

In order to test the importance of redshift upon the success of our
morphological classifier we return to our ANN (configuration 9:9:2), 
trained previously.
If aperture effects are important then we would expect to see that the
ANN will recover the galaxy morphology of relatively distant galaxies
more accurately than for nearby galaxies, particularly
in the case of Late type galaxies. For this reason the results from
the previous section were re-determined after dividing the
testing sample into two redshift bins, $z<0.05$ 
(1533 galaxies) and $z>0.05$ (761 galaxies).

The results are summarised in Table.~\ref{aper}.  Contrary to our
expectations the success of the classification is relatively
immune to the redshift being sampled.
Clearly the situation must be more complex than a simple analysis such as this 
can resolve.  

One important aspect of the spectra being used in this
analysis which may explain this situation is the substantial seeing present
at the Anglo-Australian Telescope site. This seeing acts to `smooth out' any
spectral gradients which may be present in a given galaxy.  In general
the seeing is of the order of 1.8-2.5$''$ and so can effectively
double the area being sampled by the 2dF fibre aperture.

Another major consideration in an analysis such as this is the
stability of the morphological classification with redshift - as more
distant galaxies will tend to be fainter and less extended on the sky.
In general the robustness of a morphological classification is
difficult to assess because of its subjectivity.
It has been found in previous work (Naim \etal\ 1995) that
this subjective element results in an uncertainty of the order of 2
T-Types.  However, this figure does not incorporate the uncertainties
introduced to the classification through inclination, obscuration and
other systematic uncertainties.  All of these uncertainties add to the
importance of being able to estimate morphologies in a more robust
manner such as by correlating with galaxy spectra, or indeed for neglecting
morphology altogether and simply using a spectral-based classification
(see e.g. Madgwick \& Lahav 2001).

Because the galaxy sample considered here has been restricted to
apparent magnitudes greater than $\bj=16.5$, misclassification is not
considered to be as 
significant an issue in this analysis as it might otherwise be.  
However when repeating the above
analysis using galaxies fainter than this magnitude limit a very substantial
systematic misclassification of spirals was observed.  This was to be
expected since the spiral arms of such galaxies will become more
difficult to resolve at higher redshift (where most of the faintest
galaxies will reside), particularly for galaxies
inclined to the line-of-sight.

\begin{table}
 \caption{Success rates of the 9:9:2 ANN for different redshift slices
  in the testing data set.  
  The numbers given are percentages of correctly classified galaxies
  of each specified morphological type.}
 \begin{tabular}{@{}cccc@{}}
   \hline
   Redshift & $N_{\rm tot}$ & Early & Late  \\
   \hline
     $z<0.05$ & 1533 & 72\% & 82\% \\
     $z>0.05$ & 761 & 67\% & 82\% \\
   \hline
 \end{tabular}
 \label{aper}
\end{table}

\section{Discussion}
\label{section:specclass}

Perhaps one of the most interesting aspects of this work on recovering
galaxy morphologies from their spectra, is how closely related the results 
from advanced statistical methods appear to be to the original 2dFGRS 
spectral classification, $\eta$.  In some regards
this was to be expected, since one is always inclined to relate one's
classification to galaxy morphology during its derivation.  For example
Folkes \etal\ (1999) used a training set of 26 galaxies drawn from the
Kennicutt Atlas (Kennicutt 1992) as a training set to derive the
original 2dFGRS spectral classification.  These 26 galaxies were 
`projected' onto the
($pc_1$,$pc_2$) plane defined by the 2dFGRS spectra and lines were
drawn by-hand to roughly separate the galaxies according to their
assumed morphologies.  However, in the case of $\eta$ this method was not
used, rather the galaxies were classified solely on the basis of
finding the most statistically significant projection in the PCA which
was robust to the known instrumental uncertainties in the 2dF
instrument (see Madgwick \etal\ 2002 for more details).

\begin{figure}
\epsfig{file=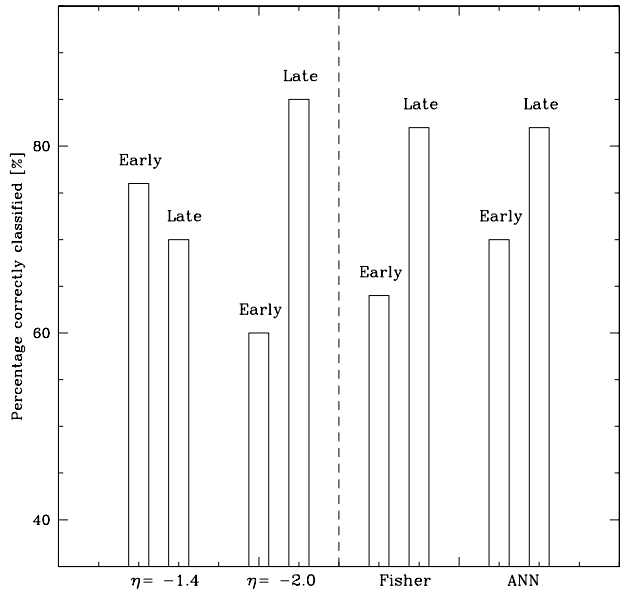,width=3.5in}
\caption{Comparison between the success rates of different classification 
methods. The first two sets of histograms show the success rates that can be 
achieved simply by using the (PCA based) $\eta$ spectral classification
adopted in the 2dFGRS.  The second two histograms show the success rates
for the more advanced statistical methods: Fisher's linear discriminant 
and the ANN.  It can be seen that the results are generally
comparable, although the ANN gives the best results.}
\label{fig:res}
\end{figure}

The overall success of correlating galaxy morphologies and spectra, using
the methods considered in this paper, is
summarised in Fig.~\ref{fig:res}, where the percentages of
galaxies classified correctly is shown for each method.  
In the case of the $\eta$ spectral
classification and the Fisher discriminant it is possible to change
the relative success rates between types simply by adopting different
`cuts' in these continuous parameters.  This is demonstrated in
Fig.~\ref{fig:res} where the success rates are shown for both an
$\eta<-1.4$ cut (2dFGRS default) and an $\eta<-2$ cut.
  
In general, the most successful method of correlating galaxy
morphology with 
spectra appears to be the Artificial Neural Network, which can achieve
consistently high success rates for both types of galaxies.  
However, in deciding
whether to implement such an algorithm the relative pay-off must be
weighed against the additional complexity of this algorithm.  This is
particularly true for this data-set since all three methods of
classification make
roughly similar distinctions between morphological types.  However, with
the advent of higher resolution (and signal-to-noise) spectra, it is
possible that such advanced methods will become much more successful.

Note that in practical situations 
the interpretation of Fig.~\ref{fig:res} is not as straight-forward  
as it might at first seem.  One must also bear in mind the relative
fractions of different galaxy types in the sample under
consideration.  In the case of the 2dFGRS, there are 
over 2 times as many Late type galaxies as Early types (2610 and 1289
galaxies respectively), so a decrease in the success rate of classifying
Late types by 5\% implies there will be an additional 250 galaxies
classified as Early type, which is at least a 10\% contamination of the
Early type sample.  This is particularly important if one wants to create a 
relatively `pure' sample of a particular type of galaxy (e.g. for efficient 
use of telescope time during observational follow-ups).

\section{Conclusions}
\label{section:conclusion}

Establishing a firm link between a galaxy's morphology and its
spectrum is advantageous for several reasons.  For instance, galaxy
spectra can be accurately determined to much greater redshifts and for
fainter objects than morphologies.  Also, most large redshift surveys
currently taking place will contain many thousands of galaxy spectra
but little information relating to the optical morphologies of
those galaxies.  In particular the separation of different
morphological types of galaxies
in these redshift surveys will be very useful as a means of
separating objects for
follow-up observations to determine independent distance
measurements using either $D_n-\sigma$ or the Tully-Fisher relation.

In this paper I have tried to quantify the link between galaxy spectra
and morphology using several advanced statistical methods; namely,
Fisher's linear discriminant and Artificial Neural Networks.
The best results produced suggest that it is possible to use
optical galaxy spectra to create galaxy samples 
containing 70\% of the Early type galaxies present and 80\% of
the Late types respectively.  
The contamination between these samples depends on the
morphological mix of the survey under consideration.  In the case of
the $\bj$-selected 2dFGRS the most significant contamination will be
of mis-classified Late types in the Early type sample ($\sim40\%$
contamination), in the case of a near-infrared selected sample
this situation will be reversed.

Essentially the results obtained using more advanced
statistical techniques
(Sections 4 and 5) are comparable to those that could be obtained
simply using the default 2dFGRS spectral classification $\eta$
(Madgwick \etal\ 2002) which can be accessed from the 2dFGRS
database\footnote{ Spectral extension parameter {\tt ETA\_TYPE}}.  This
is an interesting result and certainly adds significantly to the
physical interpretation of this parameter.

Another interesting aspect of this analysis is that the Fisher
discriminant (Section 4) identified the 4000\AA\ break to be the most
essential element of a galaxy's spectrum for the purposes of
estimating its morphology.  This result is somewhat expected since the
general correlation between galaxy morphology and colour is already
well established. 
However, it is intriguing to see this result derived in a quantitative
manner from the observed spectra themselves.

The results presented in this paper are essentially limited by the
coarseness of the morphological classification adopted, which for practical
reasons can only be divided into two separate types (rather than a more
realistic sequence of types).  As larger samples
of more accurately morphologically classified galaxies become 
available it will
be interesting to repeat the analysis presented here, in 
order to determine whether even more
information can be recovered to link a galaxy's morphology and spectrum.

\section*{Acknowledgments}

I wish to thank Ofer Lahav for suggesting this
project and providing plenty of invaluable advice and suggestions.
Raven Kaldare and Andrew Firth were very helpful in explaining
the intricacies of Artificial Neural Networks to me.  I would also
like to thank the anonymous referee for their helpful comments on the
draft of this paper.
The efforts of the 2dFGRS collaboration,
in preparing and compiling the data used in this analysis, are
greatly appreciated.  This work was supported by an Isaac
Newton studentship from the Institute of Astronomy and Trinity
College, Cambridge.

\bsp
\label{lastpage}
\end{document}